\documentclass[journal]{IEEEtran}
\usepackage{dblfloatfix}
\usepackage{xspace, amsmath, amssymb, amsfonts, epsfig, syntonly}
\usepackage{diagbox}
\usepackage{cite, bm, color, textcomp}
\usepackage{epstopdf}
\usepackage{empheq}
\usepackage{multirow}
\usepackage{comment}
\usepackage{booktabs}
\usepackage{graphicx}
\usepackage[ruled, linesnumbered]{algorithm2e}
\usepackage{verbatim}
\usepackage{threeparttable}
\usepackage[utf8]{inputenc}
\usepackage[caption=false, font=footnotesize]{subfig}
\usepackage[normalem]{ulem}
\usepackage{url}

\newcommand{\cmmnt}[1]{\ignorespaces}

\long\def\symbolfootnote[#1]#2{\begingroup \def\thefootnote{\fnsymbol{footnote}} \footnote[#1]{#2}\endgroup}

\psfull

\allowdisplaybreaks[4]

\begin{document}
\title{Goodput Maximization for Large Language Model Edge Inference: A Two-Phase
    Maskable PPO Approach\\
    }

\author{Xiaojing Chen, Qi Zhang, Wei Ni,~\IEEEmembership{Fellow, IEEE}, Shunqing Zhang,~\IEEEmembership{Senior Member, IEEE}, and Yanzan Sun}

\thispagestyle{empty}
\maketitle
\symbolfootnote[0]{$^{\dagger}$Work in the paper was supported by Shanghai Municipal Science and Technology Commission Foundation grants 25DP1500300 and 24DP1501001. 
\textit{(Corresponding author: Yanzan Sun.)}

X. Chen, Q. Zhang, S. Zhang and Y. Sun are with the Key Laboratory of Specialty Fiber Optics and Optical Access Networks, Shanghai University, Shanghai 200444, China. 
Emails:~\{jodiechen, zhangqi$\_$kawhi2, shunqing, yanzansun\}@shu.edu.cn.

W. Ni is with the School of Engineering, Edith Cowan University, Perth, WA 6027, Australia. Email: wei.ni@ieee.org.
}

\vspace{-0.37cm}

\begin{abstract}
This paper presents a novel two-phase maskable proximal policy optimization (TP-MPPO) algorithm, which maximizes the system goodput counting request throughput with strict service level objective (SLO) compliance
for large language model (LLM) inference services in wireless edge networks. In the first phase of TP-MPPO, we optimize the task offloading decisions by MPPO with action masking mechanism, effectively avoiding exploring invalid actions and reducing the action space. In the second phase, closed-form solutions are derived for uplink bandwidth allocation; a greedy algorithm is designed for downlink bandwidth allocation to provide immediate rewards for the MPPO in the next round. The two stages alternate till convergence. Simulation results demonstrate that TP-MPPO can improve the system reward by 33.3\%--87.5\% compared to its benchmarks and achieve the highest goodput.
\end{abstract}

\section{Introduction}
The rapid proliferation of large language models (LLMs) has surged inference
demands, straining traditional cloud-based LLM inference services that struggle
with throughput and latency \cite{10759588}. Edge computing supplements cloud-based LLM inference by leveraging distributed resources to scale service capacity. Furthermore, deployment on nearby edge nodes effectively  mitigates propagation latency.

Unlike conventional edge scenarios, LLM services are constrained by intensive workloads and GPU scarcity. At the model and system level, efficient inference techniques such as KV cache compression~\cite{liu2024kivi} and speculative decoding~\cite{10812936} reduce per-request compute overhead, while selective batching~\cite{yu2022orca} improves server-side utilization. These techniques are complementary to network-level resource management, which requires jointly optimizing task offloading and bandwidth allocation across heterogeneous users and wireless edge nodes~\cite{10591707,11095279}. In \cite{10591707},
    LLM task offloading, computational, bandwidth and graphics memory resource
    allocation were optimized to minimize the average latency of all LLM tasks. In
    \cite{11095279}, LLM task distribution, computational and communication resource
    allocation were optimized to maximize the system utility of 6G networks.
    However, the studies in \cite{10591707} and \cite{11095279} overlooked throughput, a critical
    metric for LLM service systems.

Although some studies have targeted aggregate throughput~\cite{10759588}, they have overlooked the fact that downstream applications have different latency requirements for user experience, leading to dramatically different service level objectives (SLOs) that need to be satisfied.
The most widely used SLOs in LLM services are Time-to-First
Token (TTFT), Time per Output Token (TPOT) and End-to-End (E2E) latency.
Goodput \cite{zhong2024distserve}, which captures request throughput under SLO
attainment reflecting both cost and service quality, can be an effective metric for
LLM service systems. 
To fill the void in goodput-oriented optimization, this paper presents a two-phase maskable proximal policy optimization (TP-MPPO) algorithm, which determines offloading and bandwidth allocation for edge LLM inference with GPU-constrained users, while preventing out-of-memory (OOM) failures caused by excessive batch sizes. The contributions of this paper are collated below.
\begin{itemize}
\item We model a wireless edge LLM inference system capturing full E2E latency pipeline and GPU memory constraints, enabling a realistic formulation of joint offloading and resource allocation.
\item We formulate a mixed-integer nonlinear programming (MINLP) problem to maximize goodput, bridging the gap between raw throughput and service quality. The problem jointly optimizes task offloading and bandwidth allocation under SLO and memory capacity constraints.
\item We propose TP-MPPO, which decomposes the MINLP into an MPPO-based offloading phase and an analytical bandwidth allocation phase with closed-form uplink and greedy downlink solutions. This algorithm provides immediate rewards that drive stable policy convergence, e.g., within 200 episodes.


\end{itemize}

Experiment results demonstrate that TP-MPPO improves the system reward by 33.3\%--87.5\% compared to its benchmarks and achieves the best goodput performance under diverse configurations of user and node numbers.

The rest of this paper is organized as follows. Section~II provides the system model. Section III formulates the considered problem. The TP-MPPO algorithm is proposed in Section IV and evaluated numerically in Section V, followed by concluding remarks in Section VI.

    \section{System Model}
    \begin{figure}[t]
        \centering
\includegraphics[width=0.4\textwidth]{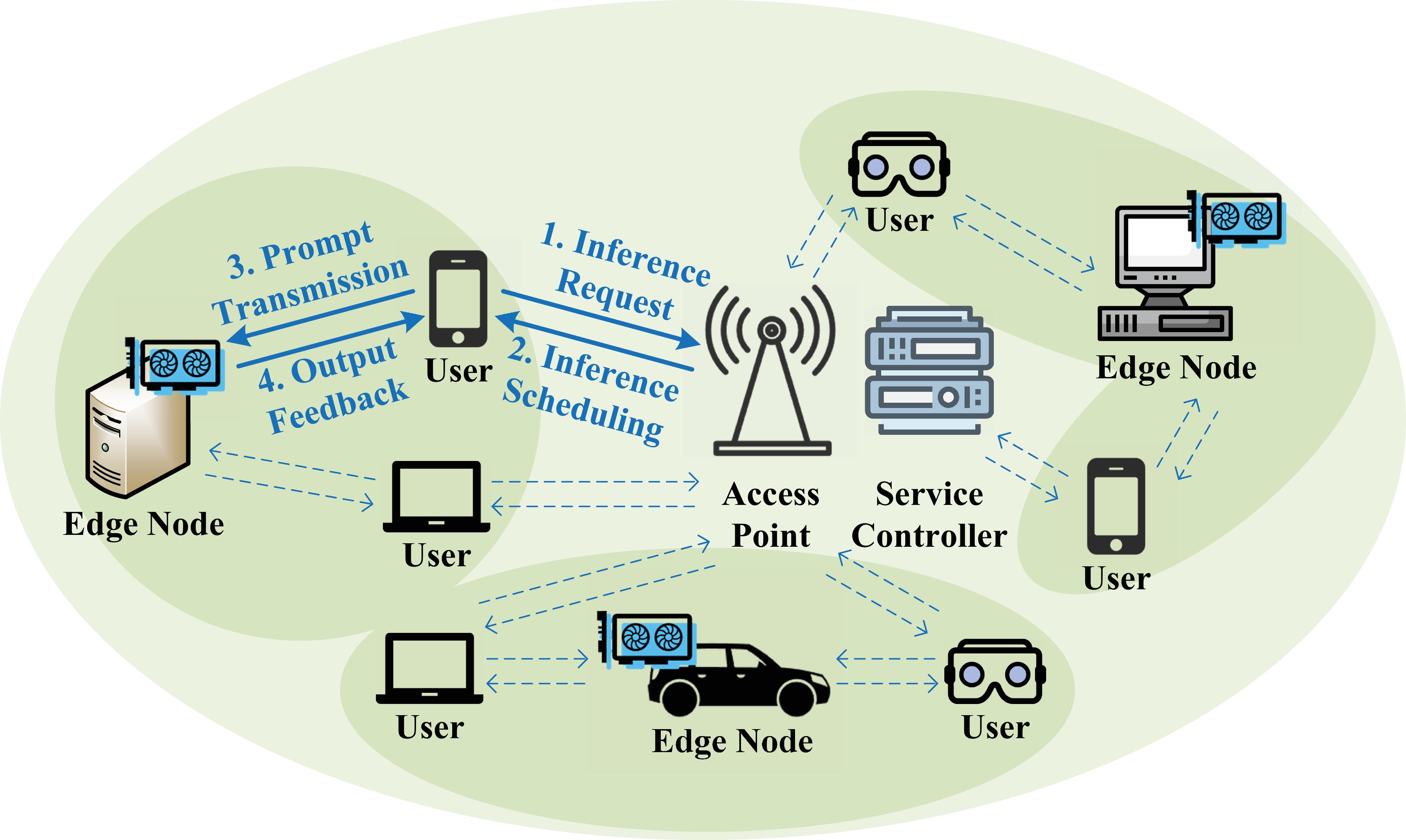}
        \caption{Illustration of the system supporting LLM inference services.}
        \vspace{-4mm}
        \label{fig1}
    \end{figure}
    We consider an LLM inference system comprising a service controller co-located with an access point (AP), $J$ GPU-enabled edge nodes ($\mathcal{J}= \{ 1,\ldots,J \}$), and $N$ users ($\mathcal{N}= \{ 1, \ldots,N \}$); see Fig.~\ref{fig1}. These users
    generate LLM inference requests but cannot perform local inference due to
    lack of GPU resources; they need to offload these tasks to the edge nodes.
    The AP acts as a centralized scheduler and controller, collecting requests
    and making offloading decisions \cite{10759588}. 

\begin{figure}[t]
        \centering
        \includegraphics[width=0.36\textwidth]{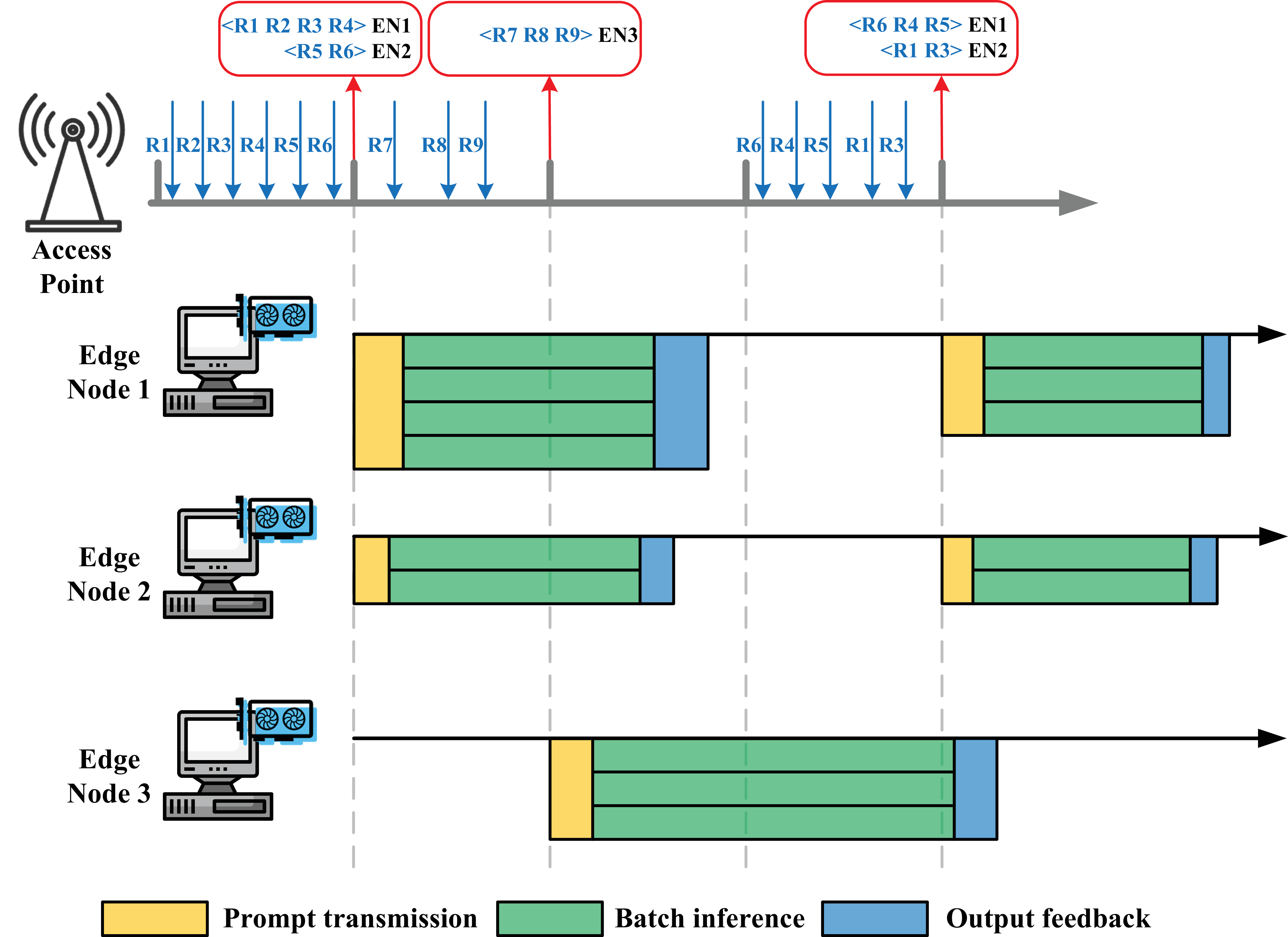}
        \caption{Workflow of LLM inference system: Requests R1--R6 arrive in the first slot; the AP assigns R1--R4 to Node 1 and R5--R6 to Node 2. Subsequently, R7--R9 are allocated to Node 3, since Nodes 1 and 2 remain occupied with ongoing inference.}
        \vspace{-3mm}
        \label{fig2}
    \end{figure}

    As depicted in Fig.~\ref{fig2}, the AP schedules requests on the basis of time slots $\mathcal{T}= \{ 1,\ldots,K \}$. {\color{black}The slot length is denoted as $\tau$, which acts as a macroscopic scheduling window rather than an orthogonal frequency division multiplexing (OFDM) symbol duration.}   Requests arriving in slot $t$ are buffered at
    the AP and scheduled at the end of this slot. Each request from
    user $n$ is denoted by $q_{n}^{t}= \left\langle s_{n}^{\mathrm{in},t}, s_{n}^{\mathrm{out},t}, \varsigma_{n}^{t}\right\rangle$,
    where $s_{n}^{\mathrm{in},t}$ is the input prompt length, $s_{n}^{\mathrm{out},t}$
    is the maximum output sequence length, and $\varsigma_{n}^{t}$ defines the E2E delay bound. The AP decides whether to offload a request to node
    $j$ or reject it. The offloading decision of request
    $q_{n}^{t}$ is an integer variable
    $\alpha_{n}^{t}\in \mathcal{J}\cup \{0\}$, and $\alpha_{n}^{t}= 0$ indicates that
    the request is rejected and failed. 
    After scheduling, users transmit their inference prompts to the designated edge nodes, which aggregate the prompts into a single batch, perform inference, and return the output results.

    We assume each user initiates a new request only after receiving the previous response \cite{yu2022orca},  with ``thinking time'' following an exponential distribution. 
The LLM inference process comprises two stages\cite{10811953}: the $\mathit{prefill~stage}$,
which processes inputs in parallel to initialize the Key-Value (KV) cache and generate
 the first token; and the $\mathit{decoding~stage}$, which utilizes the KV cache to auto-regressively generate subsequent tokens. Both stages rely on the Transformer architecture \cite{vaswani2017attention}, with the key distinction that prefill computes self-attention across all input tokens concurrently, whereas decoding restricts attention to the new token and cached history. Let $o_{n}^{t}\le s_{n}^{\mathrm{out},t}$ denote the total number of output 
tokens generated for request $q_{n}^{t}$. 

\subsection{Communication Model}
The communication model focuses on prompt transmission and output feedback,
    with negligible overhead from inference request and scheduling. 
    We consider OFDM, with each edge node operating on non-overlapping frequency bands to avoid interference. Let
    $\beta_{nj}^{\mathrm{up},t}, \beta_{nj}^{\mathrm{down},t}\in[0,1]$ denote the bandwidth fractions
    assigned by edge node $j$ to user $n$ ($\alpha_{n}^{t}= j$) for uplink and
    downlink transmissions, respectively. Per time slot $t$, the uplink transmit rate of user
    $n$ is given by
    \begin{equation}
        R_{nj}^{\mathrm{up},t}=\beta_{nj}^{\mathrm{up},t}B_{j}^{\mathrm{up}}\log_{2}
        (1+\frac{p_{nj}^{\mathrm{up},t}h_{nj}^{\mathrm{up},t}}{\psi}),
    \end{equation}
    where $B_{j}^{\mathrm{up}}$ is the uplink bandwidth of edge node $j$; $\psi$
    is the background noise power at the edge node; $p_{nj}^{\mathrm{up},t}$ denotes
    the uplink transmit power of user $n$; and $h_{nj}^{\mathrm{up},t}$ is the channel
    gain from user $n$ to edge node $j$. The wireless channel consists of large-scale path loss and 
small-scale fading: $h_{nj}^{\mathrm{up},t}=\xi_{nj}^{\mathrm{up},t}\bar{h}_{nj}^{\mathrm{up}}$, where $\xi_{nj}^{\mathrm{up},t}$ follows an exponential distribution with unit mean corresponding to
Rayleigh fading, and $\bar{h}_{nj}^{\mathrm{up}}=(A_{d}\frac{3\times10^{8}}{4\pi f_{c}d_{nj}})^{d_{e}}$ is the distance-dependent average channel gain, with $A_d$ the antenna gain, $f_c$ the carrier frequency, $d_e$ the path loss exponent, and $d_{nj}$ the distance between user $n$ and edge node $j$. The channel gain $h_{nj}^{\mathrm{up},t}$ is treated as slot-level effective average value under a quasi-static fading assumption, as widely adopted in wireless edge computing models~\cite{10759588, 10591707, 11095279}. Likewise, the downlink transmit rate
    is $R_{nj}^{\mathrm{down},t}=\beta_{nj}^{\mathrm{down},t}B_{j}^{\mathrm{down}}
    \log_{2}(1+\frac{p_{nj}^{\mathrm{down},t}h_{nj}^{\mathrm{down},t}}{\psi'})$, where $B_{j}^{\mathrm{down}}$ is the downlink bandwidth of edge node $j$; $\psi'$
    is the background noise power at user $n$; $p_{nj}^{\mathrm{down},t}$ denotes
    the downlink transmit power of edge node $j$; and $h_{nj}^{\mathrm{down},t}$ is the channel
    gain from edge node $j$ to user $n$.

    \subsection{Latency and Memory Analysis}
    For request $q_{n}^{t}$ (if user $n$ generates a request during slot $t$), let $T_{n}^{\mathrm{up},t}$ denote its uplink latency, and $D_{n}^{\mathrm{in},t}$ denote the data size of its input prompt. We have $T_{n}^{\mathrm{up},t}=D_{n}^{\mathrm{in},t}/R_{nj}^{\mathrm{up},t}$. Likewise, the downlink latency of $q_{n}^{t}$ is $T_{n}^{\mathrm{down},t}=D_{n}^{\mathrm{out},t}/R_{nj}^{\mathrm{down},t}$, where $D_{n}^{\mathrm{out},t}$ is the data size of its output. \textcolor{black}{Each uplink and downlink transmission is assumed to complete within one scheduling slot, over 
which the channel gain is treated as quasi-static. Extending the framework to account for transmissions spanning multiple slots is left for future work.}
    
    The E2E latency of user $n$'s task is composed of five components. Since requests
    arriving at the AP are scheduled at the end of each time slot, a waiting
    time for scheduling $T_{n}^{\mathrm{wait},t}$ is incurred at the AP. Upon
    finishing the prompt transmission which takes $T_{n}^{\mathrm{up},t}$, user
    $n$ may need to wait for other users assigned to the same node to complete their
    prompt transmissions before conducting batch inference, resulting in a batching
    wait time $T_{n}^{\mathrm{stay},t}$ at the edge node. Let $\mathcal{N}_{j}^{t}= \{ n|\alpha_{n}^{t}= j\}$ denote the set of requests assigned to edge node $j$, with batch size $b_{j}^{t}= |\mathcal{N}_{j}^{t}|$. During batch inference at edge node $j$, all requests assigned 
to the same batch are processed together, and the batch completion latency is given by
$T_{j}^{\mathrm{inf},t}=T_{j}^{\mathrm{P},t}+T_{j}^{\mathrm{D},t}$,
where $T_{j}^{\mathrm{P},t}$ and $T_{j}^{\mathrm{D},t}$ denote the prefill and decoding latency, respectively. Following the analytical latency model for Transformer inference 
in~\cite{10759588}, the prefill latency is $T_{j}^{\mathrm{P},t}=L b_j^t(6s_j^t h^2+(4(s_j^t)^2h+2s_j^t h^2)+4s_j^t h \mathfrak{h})/F_j$ and the decoding latency is $T_{j}^{\mathrm{D},t}=L b_j^t(o_{j}^{t}-1)(6h^2+(4(s_j^t+\frac{o_{j}^{t}}{2})h+2h^2)+4h\mathfrak{h})/F_j$. Here,  $s_j^t$ is the padded input length, $o_j^t=\max_{n\in\mathcal{N}_{j}^{t}}o_n^t$ is the batch-level output length, $L$ is the number of Transformer layers, $h$ is the Transformer hidden dimension, $\mathfrak{h}$ is the feed-forward network (FFN) hidden dimension, and $F_j$ is the computational capacity (in FLOPs/s) of edge node $j$. The inference latency model captures the latency trends 
observed in practical inference systems, where larger batch sizes, longer sequences, and larger models increase computation latency, while stronger GPU resources reduce latency.
    Finally,
    the output feedback consumes $T_{n}^{\mathrm{down},t}$. The total E2E latency
    of user $n$ is $T_{n}^{\mathrm{E2E},t}=T_{n}^{\mathrm{wait},t}+T_{n}^{\mathrm{up},t}+T_{n}^{\mathrm{stay},t}+T_{j}^{\mathrm{inf},t}+T_{n}^{\mathrm{down},t}$.

    For batch inference, the memory footprint of edge node $j$ is
    $m_{j}^{t}= m^{\mathrm{wts}} + m_{j}^{\mathrm{kv},t}$, which consists of the memory for model weights, $m^{\mathrm{wts}}$, and the KV cache footprint, $m_{j}^{\mathrm{kv},t}$. The KV cache footprint $m_{j}^{\mathrm{kv},t} = 4hLb_{j}^{t}(s_{j}^{t}+o_{j}^{t})$~\cite{10759588}.

    \section{Problem Formulation}
  Let a binary variable $\mathfrak{r}_{n}^{t}\in\{0,1\}$ denote the completion
status of request $q_{n}^{t}$, where $\mathfrak{r}_{n}^{t}=1$ indicates that
request $q_{n}^{t}$ is successfully completed, and $\mathfrak{r}_{n}^{t}=0$
indicates otherwise. Request $q_n^t$ is considered successfully completed ($\mathfrak{r}_{n}^{t}=1$) if it is assigned to an edge node (i.e., not rejected, $\alpha_n^t \neq 0$), the batch inference finishes without an OOM failure, and the output is returned to the user. The system throughput can be given as $Y=\sum_{n\in\mathcal{N}}\sum_{t\in\mathcal{T}}\mathfrak{r}_{n}^{t}$. Larger batch sizes enhance throughput; conversely, oversized batch sizes can trigger memory exhaustion, causing the inference to fail. Let $M_{j}$ denote the memory capacity of edge node $j$, which depends on its GPU resources.

We define \emph{goodput} as the number of completed requests satisfying the E2E latency SLO \cite{zhong2024distserve}.
For request $q_{n}^{t}$, let $\mathfrak{g}_{n}^{t}\in\{0,1\}$ denote whether request
$q_{n}^{t}$ contributes to the goodput. $\mathfrak{g}_{n}^{t}=1$ only if $\mathfrak{r}_{n}^{t}=1$ and $T_{n}^{\mathrm{E2E},t}\le \varsigma_{n}^{t}$. Requests that are rejected, fail due to OOM, or are completed but violate the SLO are not counted. The system goodput is defined as $Y^{'}=\sum_{n\in\mathcal{N}}\sum_{t\in\mathcal{T}}
    \mathfrak{g}_{n}^{t}$, a dimensionless metric representing the total number of SLO-compliant completed 
requests over the entire simulation horizon (i.e., all $K$ time slots).

Our goal is to maximize the system goodput by
    optimizing offloading decisions and bandwidth allocation, accounting for the unique challenges of LLM inference driven by dynamic KV cache footprint and heterogeneous SLO requirements. Let $m_{j}^{t}$ represent the memory footprint of edge node
    $j$ for batch inference. The
    optimization problem is formulated as:
    \begin{subequations}
        \label{P1}
        \begin{align}
            \mathcal{P}1:\quad\quad&\mathop{{\rm{max}}}\limits_{\left \{\substack{\text{$\bm{\alpha}^{{t}}$,$\bm{\beta}^{\mathrm{up},t}$,$\bm{\beta}^{\mathrm{down},t}$}} \right \}_{t}}  Y^{'} \\
            \text{s.t. }~~~~~& \sum_{n\in\mathcal{N}_{j}^{t}}\beta_{nj}^{\mathrm{up},t}\le 1,~  \sum_{n\in\mathcal{N}_{j}^{t}}\beta_{nj}^{\mathrm{down},t}\le 1, ~~\forall j,t, \label{C2} \\
         & m_{j}^{t}\le M_{j}, \quad \forall j,t. \label{C3}
        \end{align}
    \end{subequations}
    Here, $\bm{\alpha}^{{t}}=  \{ \alpha_{n}^{t},\forall n \}$, $\bm{\beta}^{\mathrm{up},t}
    =  \{ \beta_{nj}^{\mathrm{up},t},\forall n,j\}$, and
    $\bm{\beta}^{\mathrm{down},t}= \{ \beta_{nj}^{\mathrm{down},t},\forall n,j\}$. Constraint~\eqref{C2} limits uplink and downlink bandwidth allocation, while Constraint~\eqref{C3} ensures that each edge node's memory footprint remains within its capacity. Solving $\mathcal{P}1$ is challenging because it is an NP-hard MINLP problem involving non-convex and discontinuous binary decisions. Moreover, LLM batch inference creates strong coupling among users, where the latency and memory consumption of a batch depend on the specific
    combination of assigned tasks.

    \section{Proposed TP-MPPO Scheme}
    We present TP-MPPO
    to address $\mathcal{P}1$. In the first phase, MPPO determines the offloading decisions. In the second phase, the remaining problem is decomposed into two subproblems. Closed-form solutions are derived for uplink bandwidth allocation, while downlink bandwidth allocation is solved using a greedy-based algorithm.

    \subsection{PPO Framework with Invalid Action Masking}
    Upon inspecting $\mathcal{P}1$, optimizing $\bm{\alpha}^{{t}}$ first 
     makes the remaining problem over
    $\{\bm{\beta}^{\mathrm{up},t},\bm{\beta}^{\mathrm{down},t}\}$ more
    tractable. We propose MPPO to tackle offloading decisions $\bm{\alpha}^{{t}}$.
   MPPO extends PPO with invalid action masking~\cite{10620275}, enabling agents to avoid infeasible actions under state-dependent constraints and complex action spaces. The MPPO components are designed as follows.
    \subsubsection{\textbf{State}}
    The state space is $\boldsymbol{s}^{t}= \{ q_{n}^{t}, h_{nj}^{\mathrm{up},t},h_{nj}^{\mathrm{down},t}, I_{j}^{t}, \break \forall n,j\}$, 
    where $I_{j}^{t}\in\left \{0,1\right\}$ indicates whether edge node~$j$ is idle, taking $1$ if idle and $0$ otherwise.
    \subsubsection{\textbf{Action}}
    To prevent invalid node selection, we introduce a binary mask vector
    $\mathbf{v}_{n}^{t}=\{v_{n,j}^{t},\forall j\}$ for each user. Each element is dynamically generated based on the current environment: $v_{n,j}^{t}=1$ when request $q_n^t$ exists and node $j$ is available at slot $t$ (i.e., $I_{j}^{t}=1$); otherwise, $v_{n,j}^{t}=0$.

    During actor network forward propagation, logits of invalid actions ($v_{n,j}^{t}=0$) are set to $-\infty$. After the Softmax, their selection probability becomes zero. This Action Masking dynamically prunes the
    action space, ensuring $\bm{\alpha}^{{t}}$ is always
    executable. The resulting action space is 
    $\boldsymbol{a}^{t}=\left \{ \bm{\alpha}^{{t}}\right \}$.

    \subsubsection{\textbf{Reward}}
    The objective is to maximize the system goodput while penalizing memory
    overflow failures. The immediate reward $r^{t}$ is formulated as a normalized
    utility given by $r^{t}= \frac{\nu_{\mathrm{good}}\sum_{n\in\mathcal{N}}\mathfrak{g}
    _{n}^{t}-\nu_{\mathrm{oom} }N^{\mathrm{oom},t}}{N^{\mathrm{req},t}}$, 
    where 
    $N^{\mathrm{oom},t}$ is the number of requests that failed due to
    memory overflow, and $N^{\mathrm{req},t}$ is the total number of requests
    generated during slot $t$. $\nu_{\mathrm{good}}$ and $\nu_{\mathrm{oom}}$ are
    weight coefficients.

    MPPO selects the action $\boldsymbol{a}^{t}$ by pruning invalid choices via the mask $\mathbf{v}^{t}= \left
    \{ \mathbf{v}_{n}^{t}\right \}_{n=1}^{N}$. For a given $\bm{\alpha}^{{t}}$, the subproblems are solved to determine $\bm{\beta}^{\mathrm{up},t}$ and $\bm{\beta}^{\mathrm{down},t}$. The complete action set $\{\bm{\alpha}^{{t}}
    ,\bm{\beta}^{\mathrm{up},t},\bm{\beta}^{\mathrm{down},t}\}$ then updates the
    environment, triggering the update of $\{ I_{j}^{t}\}$ and
    the state transition from $\boldsymbol{s}^{t}$ to $\boldsymbol{s}^{t+1}$.
    
    Specifically, the behavior actor network (with parameters $\boldsymbol{\theta}_{\mathrm{old}}$) generates action
    $\boldsymbol{a}^{t}$ under the masked policy $\pi_{\boldsymbol{\theta}_{\mathrm{old}}}
    \left (\boldsymbol{a}^{t}|\boldsymbol{s}^{t},\mathbf{v}^{t}\right )$. The critic network (with parameters $\boldsymbol{\omega}$) estimates the state value  $V_{\boldsymbol{\omega}}(\boldsymbol{s}^{t})$. Collected trajectories $(
    \boldsymbol{s}^{t},\mathbf{v}^{t}, \boldsymbol{a}^{t},r^{t},\boldsymbol{s}^{t+1}
    )$ are sampled from the replay buffer to update the target
    actor network $\boldsymbol{\theta}$ by minimizing $L_{\text{CLIP}}^{t}(\boldsymbol{\theta}) \!=\!\hat{\mathbb{E}}_{t}[ \min(
        g^{t}(\boldsymbol{\theta})\hat{A}^{t}, \mathrm{clip}( g^{t}(\boldsymbol{\theta}
        ),1\!-\!\epsilon ,1\!+\!\epsilon)\hat{A}^{t})]$, 
    where
    $g^{t}(\boldsymbol{\theta})=\frac{\pi_{\boldsymbol{\theta}}\left (\boldsymbol{a}^{t}
    |\boldsymbol{s}^{t}, \mathbf{v}^{t}\right )}{\pi_{\boldsymbol{\theta}_{\mathrm{old}}}
    \left (\boldsymbol{a}^{t}|\boldsymbol{s}^{t}, \mathbf{v}^{t}\right )}$
    is the masked-policy probability ratio, and $\epsilon$ is the clipping
    factor. $\hat{A}^{t}=\sum_{i=0}^{K-t}(\gamma )^{i}\delta^{t+i}$ is the
    advantage function calculated via generalized advantage estimation
    \cite{11440117}, where $\gamma$ is the discount factor and $\delta^{t}=r^{t}+\gamma V_{\boldsymbol{\omega}}(\boldsymbol
    {s}^{t+1})-V_{\boldsymbol{\omega}}(\boldsymbol{s}^{t})$ represents the temporal-difference error. The critic network is updated by minimizing
    $L^{t}\left ( \boldsymbol{\omega}\right ) =\mathbb{E}_{t}[ \left ( \delta^{t}
    \right )^{2}]$.

\subsection{Closed-Form Solutions to Uplink Bandwidth Allocation}
    Given $\bm{\alpha}^{{t}}$, the subset of
    users offloaded to edge node $j$ is determined. 
    For user $n\in\mathcal{N}_{j}^{t}$, the prompt transmission time $T_{n}^{\mathrm{up},t}$
    and waiting time $T_{n}^{\mathrm{stay},t}$ are dictated by the slowest
    user in the batch, as batch inference requires all prompts uploaded before padding. Thus, the uplink bandwidth allocation problem
    at node $j$ is to minimize the maximum uplink latency among its assigned users, expressed as
    $T_{j}^{\mathrm{up},t}=\max_{n\in \mathcal{N}_{j}^{t}}T_{n}^{\mathrm{up},t}$.
    The subproblem of uplink bandwidth allocation is given by
    \begin{equation}
        \label{P2}
        \begin{aligned}
            \mathcal{P}2:\mathop{{\rm{min}}}\limits_{\{ \beta_{nj}^{\mathrm{up},t}\}_{n\in\mathcal{N}_{j}^{t}}}  {T_{j}^{\mathrm{up},t}} ~~\quad
            \text{s.t. }~\sum_{n\in\mathcal{N}_{j}^{t}}\beta_{nj}^{\mathrm{up},t}\le 1,~\forall j,t.
        \end{aligned}
    \end{equation}

    According to the min-max fairness principle, the optimal strategy equalizes uplink latencies
    across all users in the batch.
    Using the auxiliary variable $\mathfrak{R}_{nj}^{\mathrm{up},t}=B_{j}^{\mathrm{up}}\log_{2}(1+\frac{p_{nj}^{\mathrm{up},t}h_{nj}^{\mathrm{up},t}}{\psi}
    )$ and equating  $\frac{D_{n}^{\mathrm{in},t}}{\beta_{nj}^{\mathrm{up},t}\mathfrak{R}_{nj}^{\mathrm{up},t}}$ for all users $n\in\mathcal{N}_{j}^{t}$, the optimal allocation is $(\beta_{nj}^{\mathrm{up},t})^{*}=\frac{{D_{n}^{\mathrm{in},t}}/{\mathfrak{R}_{nj}^{\mathrm{up},t}}}{{\textstyle \sum_{i\in\mathcal{N}_{j}^{t}}} ({D_{i}^{\mathrm{in},t}}/{\mathfrak{R}_{ij}^{\mathrm{up},t}})}
        $.

    \begin{algorithm}
        [t]
         \footnotesize
        \caption{Optimal Solution to  $\mathcal{P}3$}
        \label{alg3}
       
        \BlankLine Initialize satisfied set $\mathcal{S}_{j}^{t}=\varnothing$, and
        cumulative bandwidth usage $\beta_{j,\mathrm{sum}}^{\mathrm{down},t}=0$.

        \For{each user $n\in \mathcal{N}_{j}^{t}$}{ Calculate the minimum required bandwidth fraction $\beta_{nj,\mathrm{min}}^{\mathrm{down},t}$. }
        Sort users in $\mathcal{N}_{j}^{t}$ based on $\beta_{nj,\mathrm{min}}^{\mathrm{down},t}$
        in ascending order to obtain the sorted index list $\mathcal{K}=\{ k_{1},
        \ldots,k_{|\mathcal{N}_{j}^{t}|}\}$.

        \For{$i=1$ to $|\mathcal{N}_{j}^{t}|$}{ \If{$\beta_{j,\mathrm{sum}}^{\mathrm{down},t}+\beta_{k_{i}j,\mathrm{min}}^{\mathrm{down},t}\le 1$}{ Add user $k_{i}$ to satisfied set: $\mathcal{S}_{j}^{t}\gets \mathcal{S}_{j}^{t}\cup \{k_{i}\}$.

        Allocate minimum requirement: $\beta_{k_{i}j}^{\mathrm{down},t}=\beta_{k_{i}j,\mathrm{min}}^{\mathrm{down},t}$.

        Update usage: $\beta_{j,\mathrm{sum}}^{\mathrm{down},t}\gets \beta_{j,\mathrm{sum}}^{\mathrm{down},t}+\beta_{k_{i}j,\mathrm{min}}^{\mathrm{down},t}$. } }
        Obtain the optimal ${\{ \beta_{nj}^{\mathrm{down},t}\}_{n\in\mathcal{N}_{j}^{t}}}$.

        \BlankLine
        \vspace{-2mm}
    \end{algorithm}
    \vspace{-3mm}
    \subsection{Optimal Solution to Downlink Bandwidth Allocation}
    After batch inference, user $n$ must satisfy the SLO, with downlink latency $T_{n}^{\mathrm{down},t}$ constrained by the
    residual margin $T_{n}^{\mathrm{re},t}=\varsigma_{n}^{t}-T_{n}^{\mathrm{wait},t}- T_{n}^{\mathrm{up},t}
    -T_{n}^{\mathrm{stay},t}-T_{j}^{\mathrm{inf},t}$. The downlink bandwidth allocation problem at node $j$ seeks to maximize the number of requests meeting the SLO, 
    aligning with system goodput.
    Let
    $\mathcal{N}_{j}^{\mathrm{good},t}= \{ n|n \in \mathcal{N}_{j}^{t},T_{n}^{\mathrm{down},t}
    \le T_{n}^{\mathrm{re},t}\}$
    be the set of requests contributing to goodput. The subproblem of downlink bandwidth allocation is then formulated as
    \begin{equation}
        \label{P3}
        \begin{aligned}
            \mathcal{P}3:\mathop{{\rm{max}}}\limits_{\{ \beta_{nj}^{\mathrm{down},t}\}_{n\in\mathcal{N}_{j}^{t}}} {|\mathcal{N}_{j}^{\mathrm{good},t}|} ~~\quad
            \text{s.t. }\sum_{n\in\mathcal{N}_{j}^{t}}\beta_{nj}^{\mathrm{down},t}\le 1, ~~\forall j,t.     
        \end{aligned}
    \end{equation}

    Based on the expressions of $R_{nj}^{\mathrm{down},t}$ and $T_{n}^{\mathrm{down},t}$, we calculate the minimum required bandwidth fraction (subject to
    $T_{n}^{\mathrm{down},t}\le T_{n}^{\mathrm{re},t}$) for user $n$ by setting $T_{n}^{\mathrm{down},t}=T_{n}^{\mathrm{re},t}$:
    \begin{equation}
        \beta_{nj,\mathrm{min}}^{\mathrm{down},t}\triangleq \frac{D_{n}^{\mathrm{out},t}}{T_{n}^{\mathrm{re},t}B_{j}^{\mathrm{down}}\log_{2}(1+\frac{p_{nj}^{\mathrm{down},t}h_{nj}^{\mathrm{down},t}}{\psi'})}
        .
    \end{equation}
   By introducing $\beta_{nj,\mathrm{min}}^{\mathrm{down},t}$, we can reformulate
 problem $\mathcal{P}3$ as finding the largest subset $\mathcal{S}_{j}^{t}
    \subseteq \mathcal{N}_{j}^{t}$ that satisfies the constraint $\sum_{n\in\mathcal{S}_{j}^{t}}
    \beta_{nj,\mathrm{min}}^{\mathrm{down},t}\le 1$. 
This problem has a special equal-value structure, where all selected requests contribute
equally (i.e., one unit) to the goodput.
Hence, Algorithm~1 achieves the optimal solution by prioritizing requests with the smallest bandwidth requirements, attaining zero optimality gap. The complexity of Algorithm~1
    is dominated by the sorting operation, which is $\mathcal{O}(N\log{N})$.
    
\section{Numerical Results}
   \begin{figure}[!tbp]
        \centering
        \includegraphics[scale=0.23]{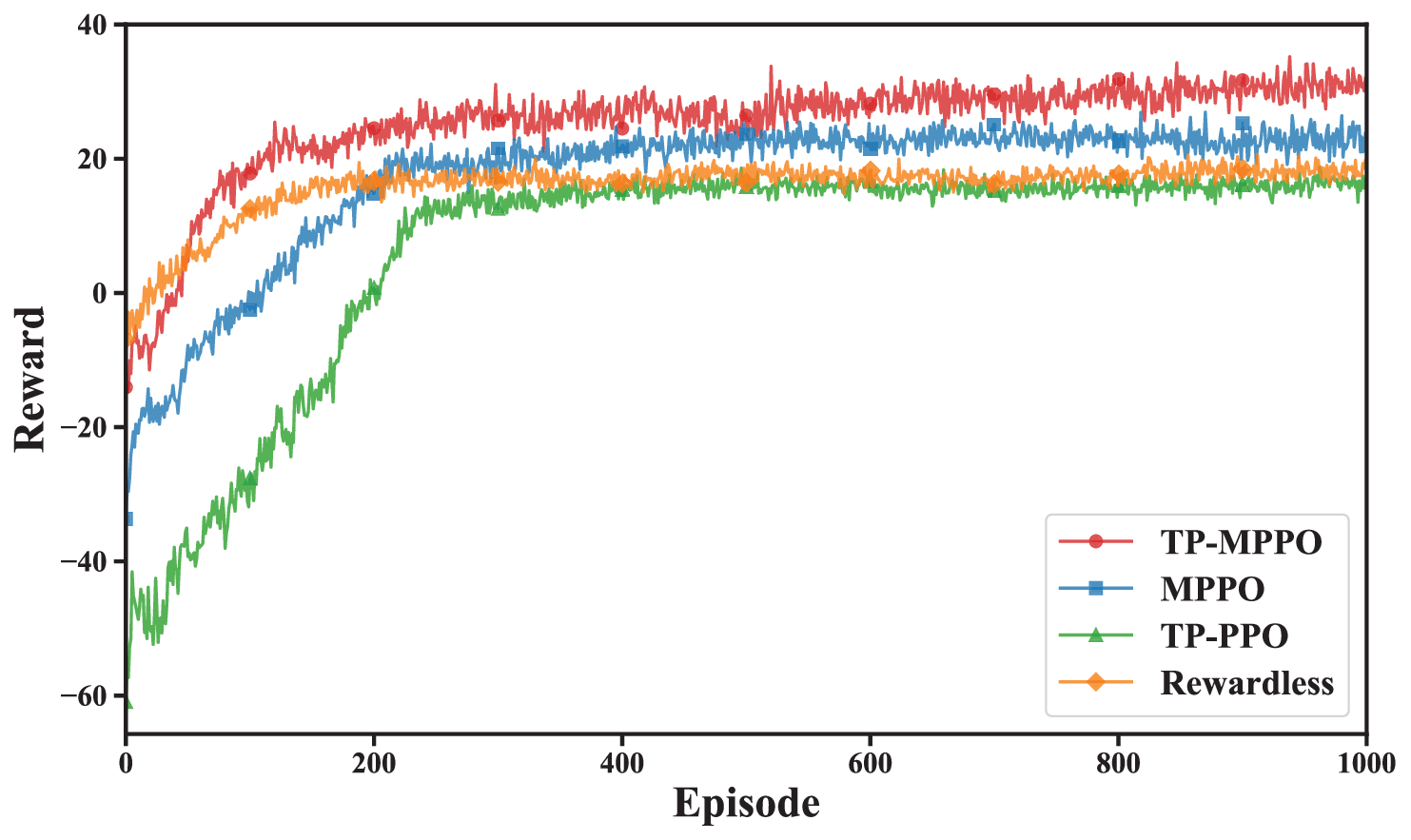}
        \vspace{-2mm}
        \caption{Reward as the number of episode grows.}
        \label{fig3}
        \vspace{-1mm}
    \end{figure}
 \begin{figure}[!tbp]
        \centering
        \includegraphics[scale=0.32]{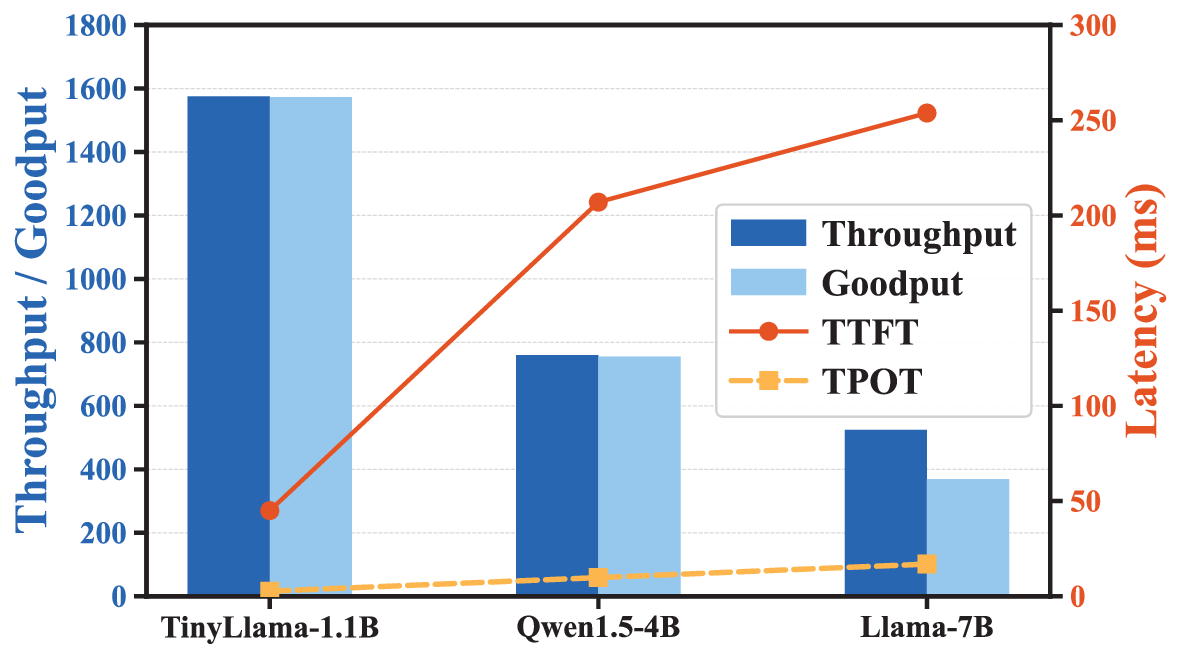}
        \vspace{-2mm}
        \caption{Comparisons of throughput, goodput, TTFT, and TPOT.}
        \vspace{-4mm}
        \label{fig4}
    \end{figure}
    
     \begin{figure*}[!t]
        \centering
        \subfloat[]{\includegraphics[width=0.25\textwidth]{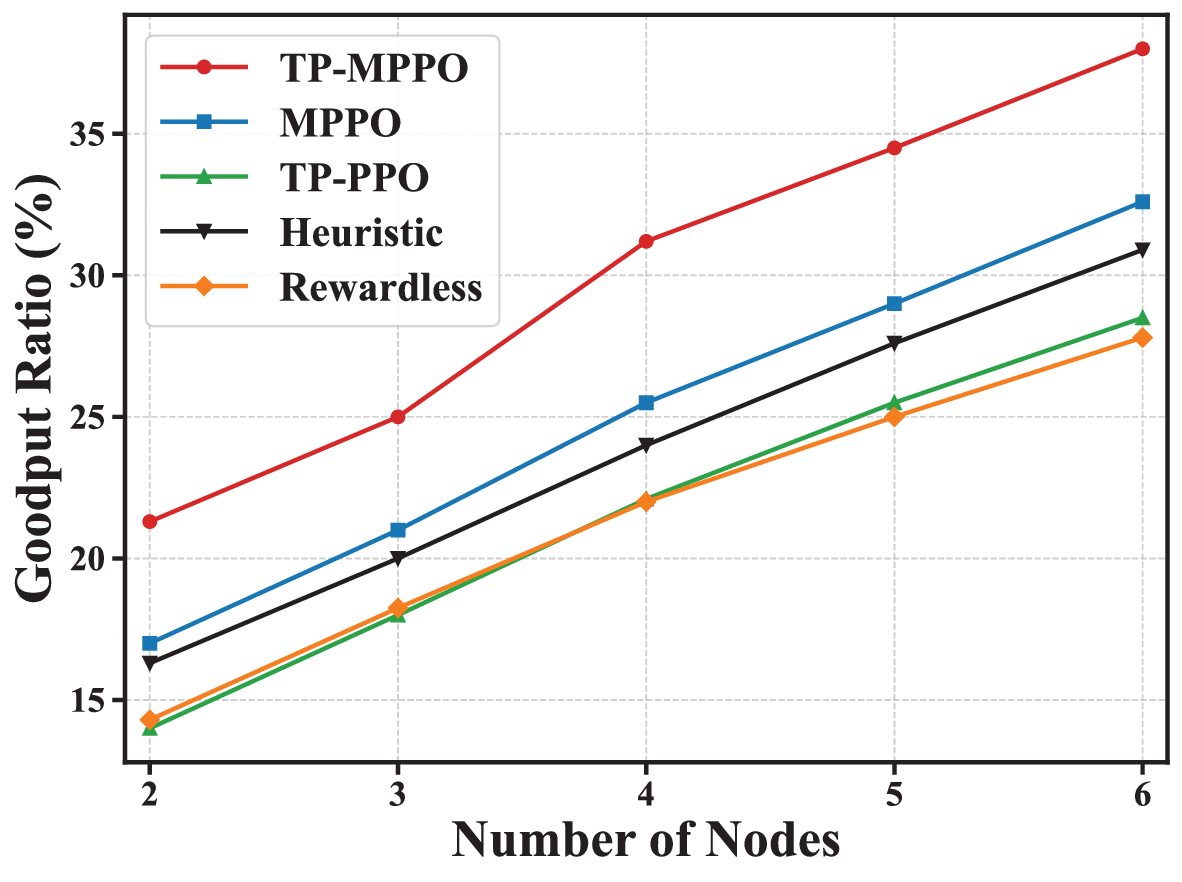}\label{fig5.1}}~~~~
        \subfloat[]{\includegraphics[width=0.25\textwidth]{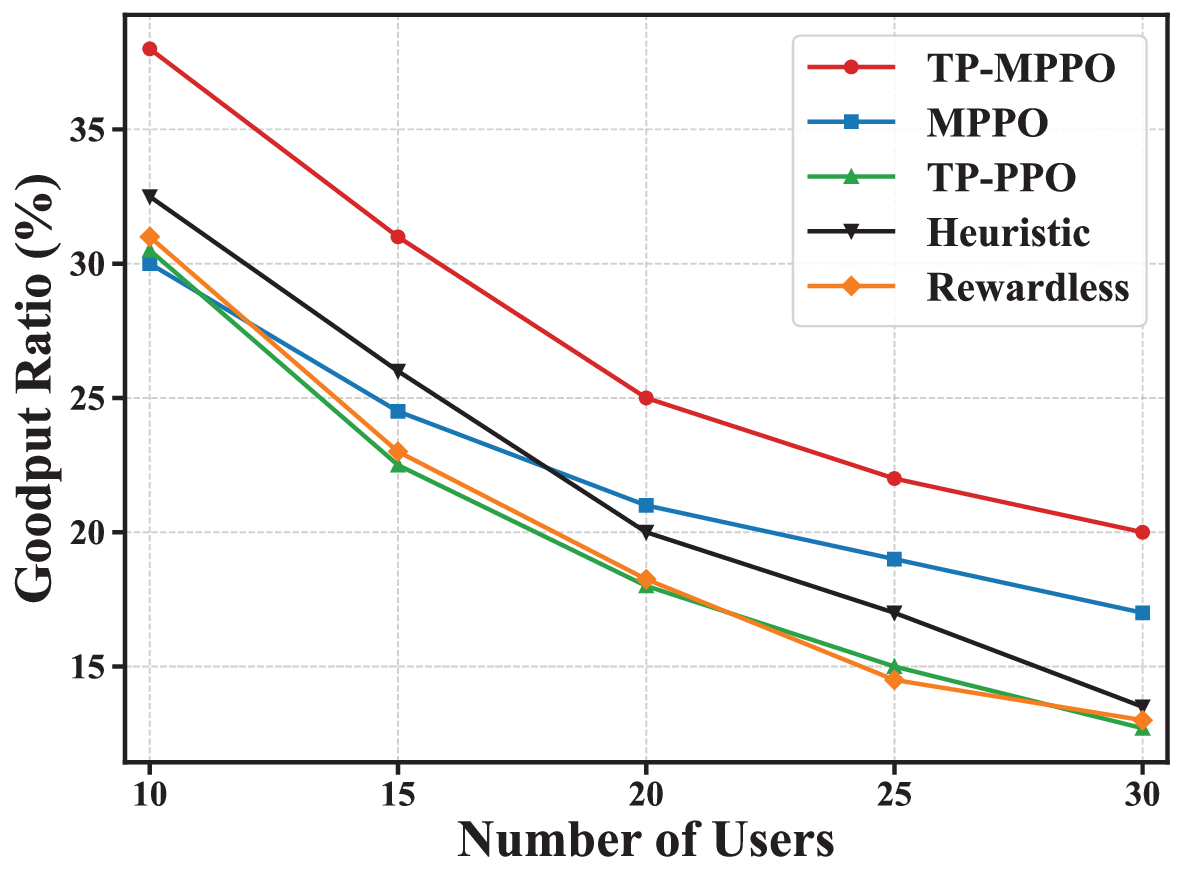}\label{fig5.2}}~~~~
        \subfloat[]{\includegraphics[width=0.25\textwidth]{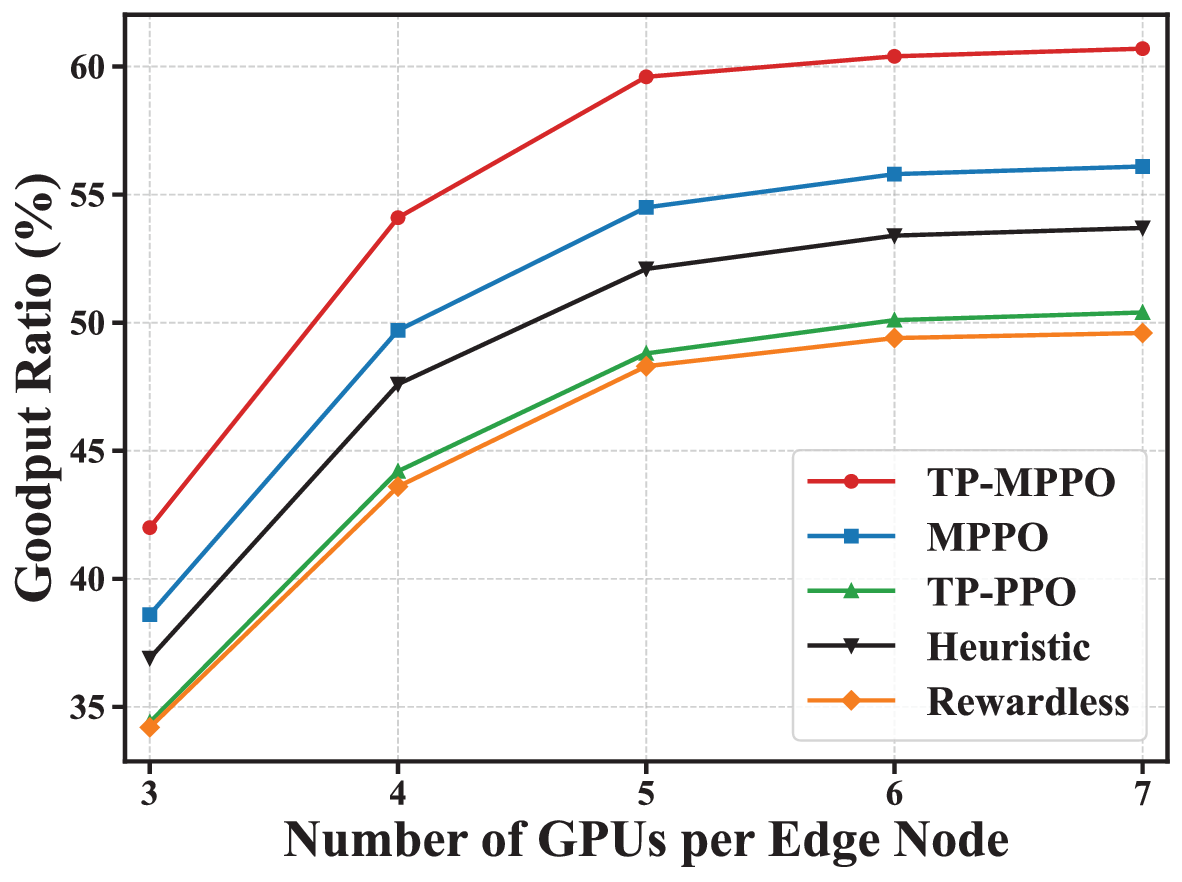}\label{fig5.3}}
        \caption{Goodput ratio with different numbers of nodes and users, and different 
 edge-node GPU provisioning levels (Llama-30B).}
        \label{fig5}
        \vspace{-0.3cm}
    \end{figure*}
    Consider a system where 3 edge nodes and 15 users are randomly deployed within a 300 m radius circular area. The input sequence length of inference request is randomly generated from $\{128,256,512\}$~tokens, with corresponding SLO targets of $\{3,5,9\}$~s. The maximum output sequence length $s_{n}^{\mathrm{out},t}$ is randomly generated
from $\{128,256,512,1024\}$~tokens, and the realized output length is set as
$o_n^t=\lambda_n^t s_{n}^{\mathrm{out},t}$, where
$\lambda_n^t\in\{0.5,0.75,1\}$~\cite{10811953,10759588}. {\color{black}All these task parameters are independently sampled 
uniformly from their respective discrete sets. 
For communication modeling, we assume 32 bits per token, consistent with the INT32 representation commonly used for token IDs.}
To ensure the simulated edge nodes reflect a realistic deployment, the memory capacity and computational capacity are set according to the specifications of an NVIDIA L2 PCIe GPU with 24~GB memory. We adopt the native Llama-7B hyperparameters: $h=4096$ and $L=32$. {\color{black}We also set $f_c=3.5$~GHz and $d_e=3$, representing sub-6~GHz fifth-generation (5G) edge access and urban/suburban non-line-of-sight (NLOS) propagation, respectively, together with $\tau=1$~s, $\psi=10^{-14}$~W, $\psi'=2\times10^{-14}$~W, $\epsilon=0.2$, $\nu_{\mathrm{good}}=1$, and $\nu_{\mathrm{oom}}=0.5$.}

    We compare TP-MPPO with four benchmarks:
    1) MPPO, which learns all variables solely via MPPO.
    2) TP-PPO \cite{10333623}, which adopts PPO without action masking for
    offloading decisions.
    3) Rewardless \cite{11095279}, which adopts classical active inference for offloading without reward guidance.
    4) Heuristic \cite{10038543}, which employs a weighted scoring mechanism
    based on channel conditions and memory load for offloading decisions.

    Fig.~\ref{fig3} shows the rewards of the considered algorithms. TP-MPPO
    yields the highest reward and converges within 200 episodes.
    The reward of TP-MPPO is 33.3\%, 87.5\% and 71.4\% higher than
    those of MPPO, TP-PPO and Rewardless, respectively. Action masking in TP-MPPO and MPPO accelerates convergence by eliminating invalid action explorations. TP-PPO struggles with early-stage cold starts, while Rewardless converges to a low reward, indicating its free-energy-based guidance is insufficient to avoid OOM states.

Fig.~\ref{fig4} shows the impact of model size on the throughput--goodput gap and latency. Lighter models achieve lower latency and higher throughput/goodput with a smaller gap, as their lower resource demands reduce SLO violations and OOM failures. In contrast, Llama-7B incurs a larger gap due to its higher memory and computational demands, highlighting the importance of goodput-oriented resource allocation and the trade-off between model capacity and service reliability.

    Figs.~5(a) and (b) demonstrate how number of edge nodes and users affect the ratio
    of goodput to total requests. In Fig.~\ref{fig5}(a) (with 20 users),  TP-MPPO surpasses all
    benchmarks. With 6 nodes, the goodput ratio of TP-MPPO
    exceeds MPPO, TP-PPO,
    Heuristic and Rewardless by 5.4\%, 9.5\%, 7.1\% and 10.2\%, respectively. In Fig.~\ref{fig5}(b) (with 3 nodes), as the user count scales to 30, the goodput
    ratio of TP-MPPO exceeds MPPO, TP-PPO, Heuristic and Rewardless by 3\%, 7.3\%, 6.5\% and 7\%, respectively.
The relatively low goodput ratios in Figs.~5(a) and (b) is caused by the constrained edge deployment scenario, where limited GPU resources and strict SLOs reduce the number of successfully served requests. Fig.~5(c) further evaluates the goodput ratio with the Llama-30B model under varying GPU provisioning levels. As GPU resources increase, TP-MPPO consistently outperforms the baselines, improving the goodput ratio from 42.0\% to 60.7\% by reducing OOM failures and inference latency.

\section{Conclusions}
    This paper presented the TP-MPPO framework for edge LLM inference services, where the task offloading decisions were learned by MPPO, uplink bandwidth allocation was given in closed form, and downlink bandwidth allocation was decided by the greedy algorithm. Simulation results have demonstrated that TP-MPPO improves the reward by 33.3\%--87.5\% and outperforms the benchmarks significantly in goodput performance across various network configurations.
 
\bibliographystyle{IEEEtran}
\bibliography{references}
\end{document}